\documentclass[twocolumn,showpacs,preprintnumbers,amsmath,amssymb,aps,superscriptaddress]{revtex4}
\usepackage{graphicx}
\usepackage{dcolumn}
\usepackage{bm}
\usepackage{mathrsfs}
\usepackage{amssymb,amsfonts}
\usepackage{natbib}

\newcommand{\bq}{\begin{eqnarray}}
\newcommand{\eq}{\end{eqnarray}}

\usepackage[dvips]{color}

\begin{document}

\title{Metastable quantum phase transitions in a periodic one-dimensional Bose gas:\\ II. Many-body theory}

\author{R. Kanamoto}
\affiliation{Division of Advanced Sciences, Ochadai Academic Production, Ochanomizu University,
Bunkyo-ku, Tokyo 112-8610 Japan}

\author{L. D. Carr}
\affiliation{Department of Physics, Colorado School of Mines, Golden, CO, 80401, USA}

\author{M. Ueda}
\affiliation{Department of Physics, University of Tokyo, Bunkyo-ku, Tokyo 113-0033 Japan}

\date{\today}

\begin{abstract}
We show that quantum solitons in the Lieb-Liniger Hamiltonian are precisely the yrast states.
We identify such solutions with Lieb's type II excitations from weak to strong interactions,
clarifying a long-standing question of the physical meaning of this excitation branch.
We demonstrate that the metastable quantum phase transition previously found in mean-field analysis
of the weakly interacting Lieb-Liniger Hamiltonian [Phys. Rev. A {\bf 79}, 063616 (2009)] extends
into the medium- to strongly interacting regime of a periodic one-dimensional Bose gas.
Our methods are exact diagonalization, finite-size Bethe ansatz, and the boson-fermion mapping
in the Tonks-Girardeau limit.
\end{abstract}
\maketitle

\section{Introduction}\label{introduction}

Exactly solvable quantum systems~\cite{94:Mattis, 97:KY} are now within reach of experiments.
This is best accomplished in highly controllable systems, such as
ultracold quantum gases~\cite{02:PS}, because one has precise control over
the effective dimensionality, so that one and two dimensions can be studied for a wide
range of interactions, both repulsive and attractive, from the weakly to the strongly interacting,
over seven orders of magnitude~\cite{09:Hulet}.
Moreover, these systems are well insulated, where there is only negligible exchange of energy
and particles with the environment, and thus suitable for the study of the
\emph{metastable quantum phase transitions} in excited states~\cite{08:CCI},
as well as ground-state quantum phases.

In this article we investigate the many-body ground and excited eigenstates
of a periodic one-dimensional (1D) Bose gas~\cite{63:LL}
under an external rotating drive going beyond the mean-field regime.
Such a geometry has been realized in experiments~\cite{05:GMMPK,06:AGR,07:M,07:NIST-ex,03:Dem}
from the weakly interacting condensate regime to the strongly interacting Tonks-Girardeau (TG)
limits~\cite{04:KWW,60:Gir,04:TG}.
In our previous analyses we showed that the average angular momentum of weakly repulsive bosons
in a one-dimensional ring undergoes a quantum phase transition (QPT) in
the {\it metastable} states as a function of interaction and rotation~\cite{08:KCU}.
In the mean-field theory this phenomenon is intuitively understood in terms of bifurcation
of stationary excited-state energy branches of the plane-wave state propagating on the ring,
and of localized soliton trains~\cite{09:KCU}.
Each excited state has a denumerably infinite number of bifurcations from the plane wave
to a state containing one or more gray or dark solitons; each such bifurcation corresponds
to a QPT.  Formally these QPTs are in fact ``crossovers'', because the two different kinds
of physical behavior, superflow and soliton, can be connected by analytic continuation.
However, these QPTs have no meaning in the thermodynamic limit, where there is
no Bose-Einstein condensation in 1D, and as such are fundamentally restricted to
the finite-size isolated systems typically found in experiments on Bose-Einstein condensates.
Moreover, such crossovers can appear quite sharp in experiments, so that the matter of terminology
becomes a question of theory, not experiment.
In metastable states of matter waves,
such as soliton trains~\cite{02:ENS-bs,02:Rice-bs}, the effects of dissipation can be suppressed
and the metastable condensate is observable.  However, this picture does not extend into
the medium- to strongly interacting regime, where quantum fluctuations cause mean-field
solitons to decay~\cite{09:MC,09:MDCC}.  Two questions follow.  (1) Does the QPT indicated by
mean-field analysis hold for stronger interactions?  (2) If so, in what way is the system
characterized on either side of this QPT, given that mean-field solitons are clearly no longer eigenstates?

Our answer lies in the special class of many-body eigensolutions called \emph{yrast states}~\cite{99:BP,99:BR}, 
defined as the lowest-energy solutions for fixed angular momentum.  Studies of one-dimensional 
systems relevant to our chosen model have a long history, including exactly solvable quantum
systems~\cite{63:LL,63:L,89:LH-2}, decay of persistent current~\cite{67:LA, 67:Litt, 61:BY, 72:PKU, 73:Blo},
and classical solitons~\cite{80:IT, 89:LH, 95:Agr}.
In the thermodynamic limit it has been known since Lieb that there exist two excitation branches
in the system, called \emph{type I} and \emph{type II} excitations.
While the physical meaning of type I was clarified as the particle excitation and
was found to agree with the Bogoliubov-type excitation in the weakly interacting limit,
the meaning of type II was elusive, described only as hole excitations~\cite{63:L}.
Seventeen years after their discovery, type II hole excitations were identified as a soliton branch
by analysis of the energy of a classical soliton in terms of the nonlinear Schr\"{o}dinger
equation~\cite{80:IT}.  However, the validity range of the nonlinear Schr\"{o}dinger equation
is limited only within the range where the matter wave possesses off-diagonal long-range coherence.

The central finding of this article is that quantum solitons in the Lieb-Liniger Hamiltonian are 
precisely the yrast states, and such states are the key to the metastable QPT previously identified 
in the mean-field context~\cite{08:KCU, 09:KCU}.
We first show how to distill the mean-field branches and QPT, which are previously found
in the mean-field theory, from the metastable yrast states. 
Throughout the manuscript, this is our basic fashion of discussing the metastable states. 
The mean-field superflow-soliton QPT found in Refs.~\cite{08:KCU, 09:KCU} is shown to be
obtained by extremizing the yrast spectra.
In the weakly interacting regime, this type II excitation can indeed be called a soliton branch
as shown in Ref.~\cite{80:IT} because of quantitative agreement
with the Gross-Pitaevskii mean-field theory~\cite{note1}.
We next introduce the concept of the ``particle'' and ``hole'' excitations which are clearly
defined in the strongly interacting Tonks-Girardeau (TG) limit.
We recover Lieb's result, and the metastable condition for the type II excitation branch
leads to the observable quantum phase with a nonintegral single-particle average momentum.
The type II ``hole'' excitation branch is made metastable (as opposed to unstable) by subjecting
the gas to a rotating drive, and is observable in typical ``rotating-bucket-type'' 
experiments~\cite{73:Legg} in a manner similar to the method used to create quantized vortices.
Finally we apply this concept to the regime of medium interaction strength.

This article is structured as follows. In Sec.~\ref{formulation} we introduce
the Lieb-Liniger Hamiltonian subject to an external rotating drive.
The yrast problem, and basic properties of the eigenstates are described.
In Sec.~\ref{diag} we investigate the many-body spectrum by exact diagonalization of the Hamiltonian
in a truncated angular-momentum basis in the weakly interacting regime, comparing with
those obtained by the mean-field theory.
In Sec.~\ref{BFmap} we study the opposite limit of the interaction strength, i.e.,
the strongly interacting TG limit where the many-body eigenproblem can be
analytically solved using the Bose-Fermi mapping.
In Sec.~\ref{BA} we address the intermediate regime of repulsive interaction between
the weakly and strongly interacting limits via the finite-size Bethe ansatz approach.
Finally, we summarize the results in Sec.~\ref{conclusion}.


\section{Formulation of the Problem}\label{formulation}

\subsection{The model and yrast states}

We consider the same model as in Refs.~\cite{08:KCU,09:KCU}, and
solve its eigenproblem beyond the mean-field and Bogoliubov theories.
The Hamiltonian for periodic one-dimensional bosons with a contact interaction,
\bq\label{LLH}
\hat{H}_0 = - \sum_{j=1}^N \frac{\partial^2}{\partial \theta_j^2}
+ g_{\rm 1D}\sum_{j<k}\delta(\theta_j-\theta_k),
\eq
is known as the Lieb-Liniger Hamiltonian (LLH)~\cite{63:LL},
where $\theta_j$ is the azimuthal angle that satisfies $0\le \theta_j < 2\pi$, $N$ the number of bosonic atoms,
and $g_{\rm 1D}$ the effective strength of $s$-wave interatomic interaction in one dimension (1D)~\cite{98:Ols}.
The length and energy units are the circumference of the ring $R$, and $\hbar^2/(2mR^2)$ with $m$ being
the atomic mass, respectively. The coupling constant is measured in units of $\hbar^2/(2mR)$
and $g_{\rm 1D}$ is hence dimensionless.
The purpose of this article is to elucidate the many-body properties of these bosons subjected to
a rotating drive.
The LLH in a rotating frame of reference with an angular frequency $2\Omega$ is given by
\bq\label{rotH}
\hat{H}(\Omega)=\hat{H}_0 - 2\Omega \hat{L} + \Omega^2N\,,
\eq
where
\bq
\hat{L}\equiv-i\sum_{j=1}^N\frac{\partial}{\partial \theta_j}
\eq
is the angular-momentum operator.
From the single-valuedness boundary condition of the many-body wave function~\cite{73:Legg},
one can show that solving the eigenproblem in the rest frame $\hat{H}_0 \Psi_0 = E(0) \Psi_0$ suffices
in order to obtain solutions to the eigenproblem $\hat{H}(\Omega)\Psi=E(\Omega)\Psi$~\cite{09:KCU}.
The eigenvalue is simply given by
$E(\Omega)=E(0)-2\Omega \langle \hat{L} \rangle +\Omega^2 N$, which is periodic with respect
to $\Omega$.

Throughout this article our approach is based on yrast problems~\cite{99:BR, 99:BP}.
Yrast, a Swedish term originally used in nuclear physics which can be translated as
``dizziest,'' refers to the lowest energy state for a given angular momentum. This approach is particularly
profitable for a finite system, because all the information about physical properties in
the rotating frame are embedded within the spectrum in the rest frame.
Thus the physical meanings of yrast states can be extracted by the simple
transformation of yrast spectra.
Since the LLH commutes with the angular-momentum operator, $[\hat{H}_0,\hat{L}]=0$, 
the yrast problem is well defined irrespective of the sign and strength of interaction,
and all the yrast states are eigenstates of both the Hamiltonians
$\hat{H}_0$ and $\hat{H}(\Omega)$.

All the eigensolutions of Hamiltonians $\hat{H}_0$ and $\hat{H}$ are classified according to the
number of atoms $N$ and total angular momentum $L$. Let us write the set of eigenstates classified into
the subspace given by parameters $(N,L)$ as $|N,L; q\rangle$ where $q \in |\mathbb{Z}|$ is
an energy quantum number that arranges the eigenvalues for fixed $(N,L)$ in ascending order.
The yrast states (the lowest-energy state under a given set of $N$ and $L$) are denoted as
$|N,L; q=1\rangle$.
The essential properties of the ground and low-lying excited states can be described
within the yrast states.  Thus we henceforth omit the quantum number $q$
from the notations for eigensolutions.
There are two external parameters, the coupling constant and the external angular
frequency of the rotating drive (divided by 2), written as $(g_{\rm 1D},\Omega)$,
respectively.
With the abbreviation of the quantum number $q=1$ and
for fixed coupling constant $g_{\rm 1D}$,
the eigenvalues that correspond to the yrast states are written as
$E_{N,L}(\Omega)$, where we explicitly write the parameter $\Omega$ in the notation
in order to clarify in which frame the system is.  With this notation, the eigensolutions
in the rest (non-rotating) frame are written as $E_{N,L}(0)$.


\subsection{Center-of-mass rotation states}\label{CMR}

Due to the translational invariance of the LLH with respect to $\theta$ and $\Omega$,
properties of a particular set of yrast states can be analyzed without solving
the problem. We denote the set of yrast states for which total angular momentum
is equal to an integral multiple of the total number of atoms $N$,
as {\it center-of-mass rotation} (CMR) states.
The energy of the CMR state takes the form
\bq\label{sf_ene}
E_{N,L=JN}(0)=J^2 N +V_{\rm int}\,,
\eq
where $V_{\rm int}$ is the interaction energy and $J \in \mathbb{Z}$ is an integer.
We call $J$ the {\it center-of-mass quantum number}, because
it physically expresses the amount of uniform translation of the center-of-mass momentum.
In the Gross-Pitaevskii mean-field theory, $J$ is conventionally called
the phase winding number; out of the mean-field regime, such terminology becomes questionable if not meaningless.
In the rotating frame, the energy of the CMR state is given by
\bq\label{sf_ene_rot}
E_{N,JN}(\Omega)=(J-\Omega)^2N+V_{\rm int}\,,
\eq
where the change in energy associated with the frame change is involved only
in the kinetic energy term, and the interaction energy
is completely separated from the parameter $\Omega$.

For repulsive interactions $g_{\rm 1D}>0$, the ground state in the absence of the
rotating drive is the state with zero angular momentum, $E_{N,L=0}(0)$.
The excitation energy of the CMR states with a finite angular momentum $L=JN$ is thus given by
\bq\label{sf_ex_ene}
E_{N,JN}(0)-E_{N,0}(0)=J^2 N,
\eq
which is independent of the strength of interaction $g_{\rm 1D}$. This is natural
because changing the total angular momentum by the amount $JN$ is just a frame change
and the interaction is isotropic.
The ground state in the presence of the rotating drive is characterized
by the CMR quantum number
\bq\label{CMqn}
J_0=\lfloor \Omega+\frac{1}{2} \rfloor,
\eq
where $\lfloor x  \rfloor$ denotes an integer that does not exceed $x$.

Because of the periodicity in the eigensolutions, an eigenstate $|N,L\rangle$ with
the energy $E_{N,L}(\Omega)$ has
a denumerably infinite number of counterparts $|N,L+JN\rangle$ and $E_{N,L+JN}(\Omega)$,
corresponding to arbitrary values of $J \in \mathbb{Z}$.
Solving the yrast problem for a limited range of fixed angular-momentum states,
e.g., $-N/2 \le L < N/2$, therefore suffices to obtain all the eigensolutions.
Moreover, the spectra are degenerate for the same magnitude of angular momentum,
$E_{N,L}=E_{N,-L}$ in the absence of rotating drive, while this degeneracy is
resolved in the presence of rotation due to the Sagnac effect~\cite{13:Sag}.
All other yrast states for $L$ out of this limited range can be obtained by
shifting the total angular momentum
by $N$ while keeping the internal structure of the eigenstates.
This is similar to a band theory concept, as discussed in Ref.~\cite{09:KCU}, with $-N/2 \le L < N/2$ playing the role of the Brillouin zone.


\section{Weakly Interacting Limit}\label{diag}

In our previous studies~\cite{08:KCU, 09:KCU}, we investigated
the weakly interacting limit of the Bose gas on a rotating ring.
The Gross-Pitaevskii equation, which corresponds to the mean-field approximation for
the Hamiltonian~(\ref{rotH}), has two kinds of solutions, namely, {\it uniform
superflow} and {\it soliton train}~\cite{00:CCR}.
The mean-field energy diagram is characterized
by the set of bifurcations of the soliton branch from the superflow branch.
These bifurcations make a continuous topological crossover in the condensate
wave function possible by changing $\Omega$.
The motivation of this section is to demonstrate how in general to distill
the mean-field branches from a sea of many-body eigenvalues.
We argue how the mean-field soliton branch, for which average angular momentum
is not quantized, emerges from the yrast spectra. The meaning of spectra related to 
symmetry breaking associated with the existence of soliton branch is
also discussed.


\subsection{Solution of the yrast problem}

To rewrite the LLH in second-quantized form, the bosonic field operator is
expanded in terms of a plane-wave basis with the single-particle
angular momentum $l$,
\bq
\hat{\psi}(\theta)=\frac{1}{\sqrt{2\pi}}\sum_{l=-\infty}^{+\infty} \hat{b}_l \,e^{il\theta}\,,
\label{eqn:discretization}
\eq
where the pre-factor of $1/\sqrt{2\pi}$ comes from the normalization of the plane wave,
and $\hat{b}_l$ and $\hat{b}^{\dagger}_l$ are annihilation and creation operators
which obey the standard
commutation relations for bosons. Equation~(\ref{eqn:discretization}) manifestly
satisfies the periodic boundary condition $\hat{\psi}(\theta)=\hat{\psi}(\theta+2\pi)$.
The Hamiltonian~(\ref{LLH}) in second quantized form is then given by
\bq\label{Ham2}
\hat{H}_0&=&\sum_{l=-\infty}^{+\infty}l^2\hat{b}_l^{\dagger}\hat{b}_l\nonumber\\
&&+g_{\rm 1D}\sum_{k,l,m,n=-\infty}^{+\infty}
\hat{b}_k^{\dagger}\hat{b}_l^{\dagger}\hat{b}_m\hat{b}_n\delta_{k+l,m+n}\,.
\eq
\begin{figure}[t]
\includegraphics[scale=0.46]{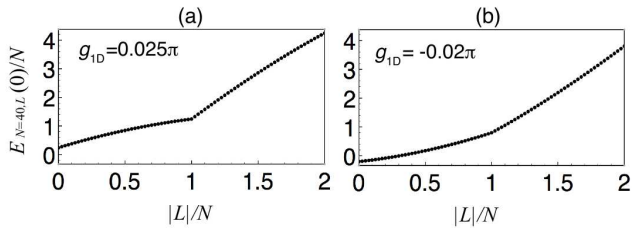}
\caption{Yrast energy eigenstates for (a) repulsive, and (b) attractive interaction
for $N=40$ bosons on the ring, obtained by diagonalization of Hamiltonian~(\ref{Ham2}) with the cutoff
angular momentum $|l_c|=2$.
The spectrum has kinks where $L$ is an integral multiple of $N$, and
is symmetric with respect to $L=0$, i.e., $E_{N,-L}(0)=E_{N,L}(0)$.
In the large $N$ limit, the number of points increases and the discrete yrast energies
approach a continuous curve while the curvature and kink points remain unchanged.
}
\label{fig1}
\end{figure}
Note that all the angular momenta are measured in units of $\hbar$.
The eigenstates can be expanded in terms of a Fock-state basis $|\{n_l\}\rangle$ that
represents the occupation number of each single-particle angular-momentum state,
\bq
|\{n_l\}\rangle=|\ldots, n_{-1}, n_0, n_1,\ldots\rangle\,.
\eq
These states satisfy the conservation laws
\bq
\sum_l n_l = N,\quad \sum_l ln_l = L\,.
\eq
In practice, for numerical calculations we require a cutoff angular momentum
$l_{\rm c}\geq 0$.
The range of the possible total angular momenta for numerical diagonalization is
hence limited to the interval of integer values $L \in [-l_{\rm c} N,l_{\rm c}N]$.
In the weakly repulsive interacting regime $g_{\rm 1D}N \lesssim \mathcal{O}(1)$,
which we study in this section, a cutoff of $l_{\rm c}=2$ provides a quantitative
agreement in energy eigenvalues~\cite{note2}
with those obtained by the Bethe ansatz shown in Sec.~\ref{BA}.
Thus all the results from this section are obtained with a cutoff of $l_{\rm c}=2$.

Figure~\ref{fig1} shows the yrast energies
$E_{N,L}(\Omega=0)=\langle N,L|\hat{H}_0|N,L \rangle$, namely,
the smallest eigenvalue obtained by the diagonalization of the Hamiltonian $\hat{H}_0$
within the restricted Hilbert space $|N,L\rangle$ for (a) $g_{\rm 1D}=2.5\times 10^{-2}\pi$ 
and (b) $g_{\rm 1D}=-2.0\times 10^{-2}\pi$ for $N=40$.
The ratio of the mean-field interaction energy to the kinetic energy
corresponding to these values of $g_{\rm 1D}$ and $N$ is
(a) $g_{\rm 1D}N/(2\pi)=0.5$, and (b) $-0.4$, respectively.
The case of attractive interaction is shown just for reference, as our main interest
is in the case of repulsive interactions.
As the cutoff angular momentum is $l_c=2$, $4N+1$ yrast states
(eigenstates corresponding to
the eigenvalues for $L\in \{0,\pm 1,\dots \pm 2N\}$) are plotted.
Recall that energies
depend only on the magnitude of the angular momentum: $E_{N,L}=E_{N,-L}$ for $\Omega=0$.

The key feature of the spectrum is that there appears a prominent kink at every CMR state $L=JN$.
As shown in Sec.~\ref{CMR}, the excitation energy of these states is given by $J^2N$.
We note, however, that other states $L\ne JN$ as well as the curvature of the spectrum
are also important and determine the existence of another quantum phase, as we show next.


\subsection{Superflow and soliton components}

\begin{figure}[t]
\includegraphics[scale=0.45]{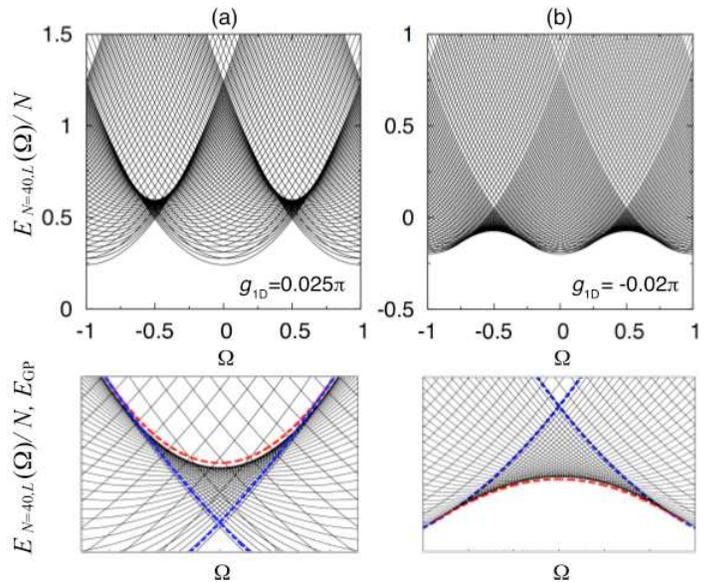}
\caption{
(Color online) Energy eigenvalues of the Hamiltonian~(\ref{rotH}) obtained by
the Legendre transformations~(\ref{Lege}) of yrast eigenstates in Fig.~\ref{fig1}.
Each curve is distinguished by a different total angular momentum $L$.
The lower panels are enlargements of the upper panels near one of
the swallowtail regions.
The dashed curves show a comparison to the Gross-Pitaevskii superflow (blue) and soliton (red) branches.
}
\label{fig2}
\end{figure}

In order to obtain the eigenstates of the LLH in the rotating frame,
we transform the yrast spectrum according to the Legendre transformation,
\bq
E_{N,L}(0)\to E_{N,L}(\Omega)=E_{N,L}(0)-2\Omega L + \Omega^2N.
\label{Lege}
\eq
Figure~\ref{fig2} plots energies $E_{N,L}(\Omega)$ as a function of $\Omega$, where
the finite number of dots, each of which is characterized by the different
angular momentum $L$ in Fig.~\ref{fig1}, become convex downward curves in Fig.~\ref{fig2}.
Each curve is thus characterized by a different total angular momentum and has a minimum
at a certain value of $\Omega$.
The degeneracy $E_{N,L}(0)=E_{N,-L}(0)$ in the absence of a rotating drive is
resolved for finite $\Omega$ due to the Sagnac effect~\cite{13:Sag}, i.e., the energy
difference naturally arises in the corotating, and counter-rotating states with
the external rotating drive.

\begin{figure}[t]
\includegraphics[scale=0.45]{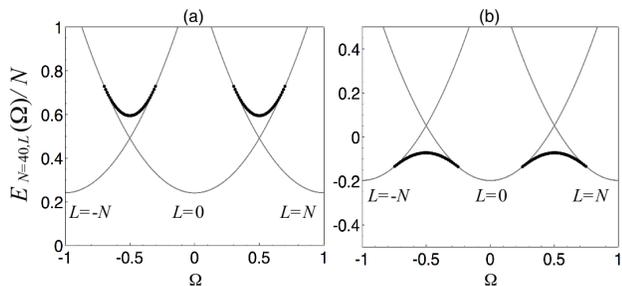}
\caption{
Energy eigenvalues of the Hamiltonian~(\ref{rotH}) that satisfy the extremization condition
$dE_{L,N}(\Omega)/dL=0$ for $N=40$ and (a) $g_{\rm 1D}=0.025 \pi$ and (b) $g_{\rm 1D}=-0.02\pi$.
Each thin curve is characterized by an integral average angular momenta,
$L/N= J \in \{0,\pm 1,\dots\}$, while
each thick point by different nonintegral angular momentum $L/N \ne J$ so that
each thick curve smoothly connects two thin curves with different angular-momentum states.
The energy of the former group agrees with the mean-field energy of the uniform superflow state,
while the latter is well approximated by the mean-field energy of solitons.
}
\label{fig3}
\end{figure}

For repulsive interactions [Fig.~\ref{fig2}(a)],
the energy $E_{N,L=J_0N}(\Omega)$ corresponds to the ground state where
$J_0$ is the ground-state CMR quantum number given by Eq.~(\ref{CMqn}).
The angular-momentum states with $L=JN$ correspond to
the CMR states, and the center of the parabola is located at
$\Omega\in \{ \mathbb{Z} \}$ at which the CMR state becomes the ground state.
On the other hand, for attractive interactions [Fig.~\ref{fig2} (b)]
the CMR state is not always a ground state, and is partially substituted for
by the nonintegral average angular-momentum states.

The transformation of the yrast spectrum according to Eq.~(\ref{Lege}) tells us that
the eigensolutions of the Hamiltonian in the rotating frame have
an extremely high density of states around $\Omega\in \{\pm 0.5, \pm 1.5,\dots\}$
due to the crossing of many eigenvalues.
These regions are enlarged in the lower panels of Fig.~\ref{fig2} for both
signs of $g_{\rm 1D}$, where we also plot the energies of the stationary states
given by the mean-field theory.
Swallowtails were previously found within the mean-field theory to
occur past the phase transition boundary between the uniform superflow and
broken-symmetry soliton states~\cite{09:KCU}.
In the microscopic quantum theory, the high-density region also forms an
upward/downward swallowtail-shape domain for repulsive/attractive interactions,
and the region is almost filled by various energy eigenvalues of
various angular-momentum states crossing each other.
The domain with the high-density swallow-tail shape looks as if it is
enclosed by the two kinds of stationary branches predicted by the mean-field theory.

\begin{figure}[t]
\includegraphics[scale=0.50]{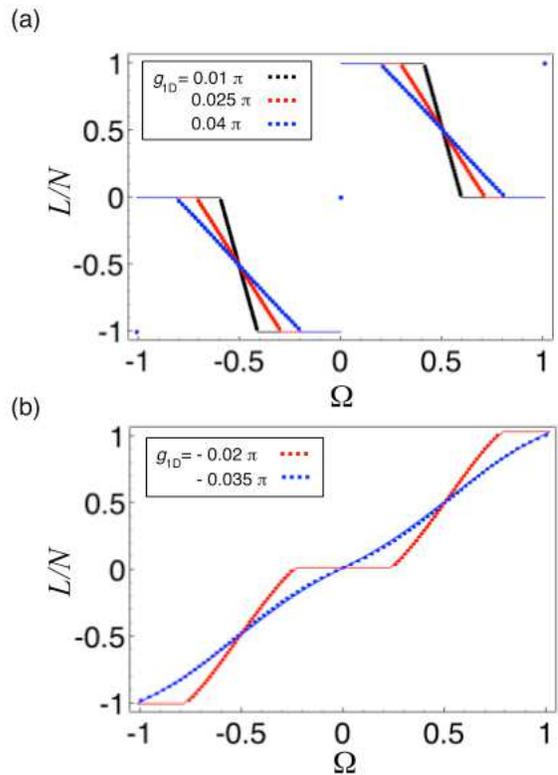}
\caption{
(Color online) Average angular momentum $L/N$ that gives $dE_{L,N}(\Omega)/dL=0$ for (a) repulsive
and (b) attractive interactions with $N=40$. Solid curves plot the single-particle
angular momentum obtained by the mean-field theory.
}
\label{fig4}
\end{figure}

Although all the angular-momentum states shown in Fig.~\ref{fig2} are eigenvalues of
the Hamiltonian~(\ref{rotH}), not all states are realized in practice.
One example is vortex formation in a scalar condensate under rotation.
Solving the yrast problem in two dimensions results in all the angular-momentum states,
including the rest condensate ($L=0$), off-axis vortex ($0 < L < N$),
a centered vortex ($L=N$), and
vortex lattices ($L > N$). In experiment, however, one drives the system with a
specific angular frequency. In such a situation, there exists a small distortion
in the trap, which ``selects'' a metastable angular-momentum state with respect to
the variation in the angular momentum of the condensate. As a result, in reality
one does not observe a stationary off-centered vortex except as a transient state.

The same argument applies to our case.
In the presence of any kind of noise, such as an infinitesimal distortion of
the trapping potential, quantum measurement of the matter wave, or whatever else
breaks the translation symmetry of the ring trap,
the {\it realizable} stationary state or {\it metastable} stationary
state is determined by extremization with respect to variations in angular momentum.
In order to find the metastable states we impose the condition
\bq\label{derE}
\frac{\partial E_{N,L}(\Omega)}{\partial L}=0
\eq
with $\Omega$ and $g_{\rm 1D}$ being fixed.

Figure~\ref{fig3} plots energy eigenvalues that satisfy the condition~(\ref{derE})
as a function of $\Omega$; and Fig.~\ref{fig4} shows the corresponding angular momentum.
These figures are quite similar to those given by mean-field theory, i.e.,
by imposing the stationary condition~(\ref{derE}) for the manifold of eigenvalues
we identify the mean-field stationary branches.
The resultant branches are classified into two kinds according to the value of the angular momentum 
and have physical meanings as follows:

{\it Superflow}: Due to the kink in the yrast spectrum at $L=JN$ in Fig.~\ref{fig1},
these CMR states always satisfy the condition~(\ref{derE}).
In particular, for the weakly interacting regime the CMR states can be specifically
called uniform superflow states, of which energies are given by Eq.~(\ref{sf_ene_rot})
and which correspond to the thin parabolic curves in Fig.~\ref{fig3}.
The energy of superflow states $E_{N,JN}$ agrees very well with the plane-wave
energy $N [(\Omega-J)^2+g_{\rm 1D}N/(2\pi) ]$ in the mean-field theory~\cite{09:KCU}.

{\it Soliton Components}: Other kinds of metastable angular-momentum states appear 
that connect distinct superflow states as a function of $\Omega$, as shown by the thick
curves in Fig.~\ref{fig3}.
The corresponding angular momentum divided by $N$ is nonintegral (see Fig.~\ref{fig4}),
but it approaches integral values at both ends of this branch.
These branches are equivalent to the maximum/minimum envelope of the high-density
swallow-tail domain for repulsive/attractive interactions, and can be approximated
by soliton energies given by the Gross-Pitaevskii mean-field theory.

However, we should {\it not} call the thick curve a soliton
branch in a rigorous sense, because each point of this branch in Fig.~\ref{fig3} is
the eigenvalue of the Hamiltonian and thus still possesses translational symmetry,
unlike solitons.
Instead we should call all the angular-momentum states inside the swallow tail in
Fig.~\ref{fig2} the {\it soliton components}, because in the presence of
infinitesimal noise these states do form a broken-symmetry state, which we denote $|\chi\rangle$.
Soliton solutions of a Gross-Pitaevskii equation can be interpreted in terms
of the eigensolutions of the many-body Hamiltonian as a state where the several
eigenvalues in the swallow-tail region are collectively superimposed.
The delocalization of a mean-field-like soliton in weakly interacting theories,
as demonstrated by Dziarmaga {\it et al.}~\cite{03:DKS}, is a dynamical
demonstration of this idea.

The energy associated with this superposition does not change significantly 
because the energy required to make it is on the order of $1/N$. As a result, 
the energy of the broken-symmetry soliton state $|\chi\rangle$ is also well 
approximated by the thick curve in Fig.~\ref{fig3}.
In the presence of an infinitesimal symmetry-breaking potential, the angular momentum
is no longer a good quantum number. However, the expectation value of the angular momentum
$\langle \chi |\hat{L}|\chi \rangle$ agrees well with that of the solitons obtained by mean-field theory, and thus behaves like that shown in Fig.~\ref{fig4}~\cite{08:KCU}.
With all these caveats in mind, we briefly say the branch drawn by thick curve
in~\ref{fig4} is the {\it quantum soliton} branch in the weakly interacting regime.

We also calculate the second derivative $d^2 E_{N,L}(\Omega)/dL^2$ with respect to
$\Omega$ in order to check whether the metastable angular-momentum state
is a local maximum or minimum.
For repulsive interactions the superflow state with a CM quantum number
$J_0=\lfloor \Omega+1/2 \rfloor$ is indeed the ground state because
the second derivative is positive at that point, while the thick points
are local maxima with respect to $L$, since the second derivative is negative.
For attractive interactions, the ground state is either a plane wave {\it or}
a bright soliton in $-0.5 \lesssim g_{\rm 1D}N/(2\pi) < 0$ but the soliton becomes
the sole ground state for $g_{\rm 1D}N/(2\pi) \lesssim -0.5$.
Consistently, the thick curve in Fig.~\ref{fig3} (b) becomes the global minimum.

To sum up this section, we obtained the yrast states $|N,L\rangle$ of the LLH
by diagonalization of the Hamiltonian in the weakly interacting regime.
Among these eigenvalues of the Hamiltonian in the rotating frame, we distilled
the metastable branches from the variety of yrast spectra by imposing an extremization
condition.
Two kinds of metastable branches, superflow and quantum soliton, were found,
consistent with mean-field theory.
The region where the different angular-momentum states in the quantum theory densely
cross indeed agrees with the soliton regime predicted by the mean-field and Bogoliubov theories.
The phrase ``quantum flesh sewn onto classical bones''
has been used elsewhere~\cite{davisED2004,bokulich2008} as a visual metaphor, perhaps
inspired by x-ray images, to describe this accord.
As we show later, the simple method shown in this section for obtaining metastable states is applicable 
to other regimes; these two branches continuously exist over a wide range of interaction,
from the weakly interacting regime all the way to the strongly interacting TG gas.


\section{Tonks-Girardeau Limit}\label{BFmap}

In the previous section, we studied the weakly interacting limit of the LLH
in the rotating frame to demonstrate how to obtain the mean-field-like stationary states.
In the opposite limit of the strongly interacting TG regime, where the bosons are impenetrable
and hence behave like spinless fermions, the eigenproblem can be calculated via
the Bose-Fermi mapping. In this section we thus solve the yrast eigenproblem of free
spinless fermions. In particular, we introduce the particle and hole excitations, which
are well defined in the fermionized gas, and show that these excitations are related
to the mean-field stationary states in the opposite weakly interacting limit.

\begin{figure}[b]
\includegraphics[scale=0.37]{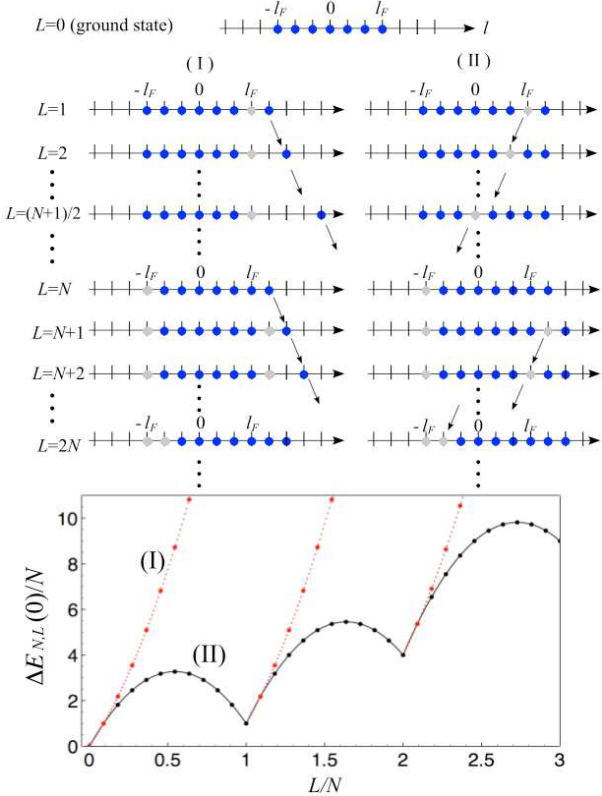}
\caption{
(Color online) Low-lying yrast ground and excited states of free fermions.
(I) Particle excitations where the angular momentum of a particle increases
while a hole is positioned at $l_F+J$.
(II) Hole excitations where a particle is placed at the lowest unoccupied
state and the angular momentum of a hole decreases.
When $L$ is an integral multiple of $N$, only the center-of-mass angular momentum
is shifted from the ground-state configuration.
The lower panel shows type I and type II excitation energies as a function
of $L/N$ for $N=11$ free fermions.
}
\label{fig5}
\end{figure}

\subsection{Bose-Fermi mapping}

The Bose-Fermi mapping theorem~\cite{60:Gir} states that the eigenvalues $E_B$ of
impenetrable bosons are identical to those of spinless free fermions $E_F$ of the same form of Hamiltonian, and the eigenfunctions of bosons $\Psi_B$ are generally written in terms of those of free fermions $\Psi_F$ as 
\bq
\Psi_B(\{\theta\})=\Psi_F(\{\theta\})\prod_{k>l}{\rm sgn}(\theta_k-\theta_l), 
\eq
where $\prod_{k>l}{\rm sgn}(\theta_k-\theta_l)$ is a unit antisymmetric function that takes the value $+1$ or $-1$ depending on the order of coordinates. This theorem holds for all the eigensolutions, and hence significantly simplifies our eigenproblem. The detailed properties of the TG gas are reviewed in Ref.~\cite{02:GW}.

We first calculate the ground- and excited-state energies of free fermions
without taking the thermodynamic limit.
For simplicity of notation we show the analytic expression only for an odd total
number of particles.
For an even number of particles the periodic boundary condition must be
taken as antisymmetric.

The ground state of $N$ (odd) free fermions is obtained by the occupation of
the lowest angular-momentum states from $l=-l_F$ to $l=l_F$ [see $L=0$ in Fig.~\ref{fig5}],
where $l_F\equiv (N-1)/2$ is the Fermi momentum. The ground-state energy is thus
\bq\label{TG_gs}
E_{N,L=0}(\Omega=0)=\sum_{l=-l_F}^{l_F} l^2 =\frac{1}{12}N(N^2-1).
\eq
We note that the $N$ dependence of Eq.~(\ref{TG_gs}) is the same
as that of the bound state for attractive interactions, $E \propto -g_{\rm 1D}N(N^2-1)$,
except for the prefactors~\cite{64:MG}.


\subsection{Particle and hole excitations}

We next consider the low-lying excitations.
Lieb has shown~\cite{63:L} that excitation of the
repulsively interacting Bose gas in the thermodynamic limit has two branches.
The first branch is called type I 
and was shown to be in agreement with the Bogoliubov spectrum of plane waves in
the weakly interacting regime.
The second branch is called type II, and this was supposed
to be absent in the Bogoliubov spectrum.
Intuitively, the type I and II branches correspond to the particle
and hole excitations, respectively.

We reconsider these branches in the context of yrast states.
For the excited state $E_{N,L}(0)$ with total angular momentum $L\ (\ne 0)$,
there exist two kinds of excitations, type I (particle excitations) and II (hole excitations), as originally named by Lieb.
To obtain these excitations one uses the following procedure. 

(I) Remove a particle at the Fermi momentum $l_F$ and
place it at the momentum $l_F+L$.
For free fermions, there is no energy-level reconstruction in an
$(N\pm 1)$-particle system associated with removal or addition of a particle.
The energy of the type I excited state $E_{N,L}^{\rm (I)}(0)$ is
thus obtained as
\bq E_{N,L}^{\rm (I)}(0)=E_{N,0}(0)-l_F^2+(l_F+L)^2.
\eq
Relative to the ground state the energy is
\bq\label{exI}
\Delta E_{N,L}^{\rm (I)}(0) \equiv E_{N,L}^{\rm (I)}(0)-E_{N,0}(0)=L(N+L-1).
\eq
There is no limitation on the single-particle angular momentum for this excitation.
Such excitations are doubly degenerate for $\Omega=0$, for $l \to -l$.

This type of excitation has an infinite set for the different CMR states $J$.
The particle excitation of the $L=JN$ state is achieved by removing
a particle at $l_F+J$ and replacing it at $l=l_F+J+L-JN$.  The resulting excitation energy is given by
\begin{eqnarray}
\Delta E_{N,L}^{\rm (I)}(0)&=&J^2N+(l_F+J+L-JN)^2-(l_F+J)^2,\nonumber\\
L &\ge& JN\,.
\end{eqnarray}

(II) Starting from the ground state, remove a particle (create a hole)
at the momentum $l_F-L+1$ and place the particle at $l_F+1$, where
$0 \le L \le N$.  It is clear from Fig.~\ref{fig5} that the hole with this kind of
low-lying excitation energy be created only within the range $-l_F \le l \le l_F$.
The energy of this excited state is given by
\bq
E_{N,L}^{\rm (II)}(0)=E_{N,L}(0)-(l_F-L+1)^2+(l_F+1)^2\,,
\eq
and
the excitation energy is thus
\bq
\Delta E_{N,L}^{\rm (II)}(0)=L(N-L+1),\qquad 0 < L \le N\,.
\eq

At $L=N$, the particle configuration in the angular-momentum space is
the same as in the ground state, provided that the center-of-mass angular
momentum is shifted. Starting from the state $L=N$, we can consider
the same kind of hole excitation, where we now place a particle at
the lowest-unoccupied angular momentum $l_F+2$, and place a hole at $l_F+2-L+N$.
In this way, the type II excitation is extended for $JN < L \le (J+1)N$ states.
This is a hole excitation of angular momentum $l_F-L+JN+J+1$ with a particle
fixed at $l_F+J$, starting from the yrast state $L=JN$.
Since the excitation energy of the CMR state $L=JN$ is $\Delta E_{N,L=JN}(0)=J^2N$,
the type II hole excitation energy is
\bq
\label{TG_exc}
\Delta E_{N,L}^{\rm (II)}(0) &\equiv& E_{N,L}^{\rm (II)}(0)-E_{N,0}(0)\nonumber\\
&=&J^2 N + (l_F+J+1)^2\nonumber\\
&&-(l_F-L+JN+J+1)^2,
\eq
where $JN < L \le (J+1)N$.

Excitation energies of type I and II are plotted in Fig.~\ref{fig5}
for $N=11$ for $L\in\{0,1,\ldots,3N\}$, i.e., up to $J=2$. The figure shows the spectra as continuous curves according to Eqs.~(\ref{exI}) and (\ref{TG_exc});
in fact the finite-size system discretizes these curves.
We note that the region $0 \le L/N \le 0.5$ was presented in Ref.~\cite{63:L}.


\subsection{Metastable hole excitation under rotation}

Next we consider the excitations under rotation, i.e., rotate all the yrast
spectra according to $E_{N,L}(\Omega)=E_{N,L}(0)-2\Omega L + \Omega^2 N$.
The energy of the type II excited state $\tilde{E}_{N,L}^{\rm (II)}(\Omega)$ for $JN < L \le (J+1)N$ measured relative to $E_{N,L=0}(0)$ is given by
\bq
\tilde{E}_{N,L}^{\rm (II)}(\Omega)=\Delta E_{N,L}^{\rm (II)}(0)-2\Omega L + \Omega^2 N\,.
\eq

\begin{figure}[t]
\includegraphics[scale=0.42]{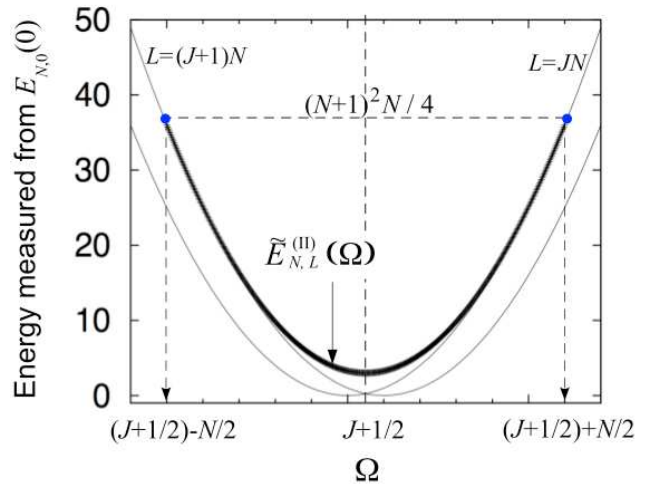}
\caption{
Realizable yrast eigenstates in the strongly interacting TG limit for $N=11$.
All the energies are plotted relative to the ground-state energy of $E_{N,L=0}(0)$.
The thin curves plot the energy of the states with angular momentum $L=JN$
and $L=(J+1)N$, while the bold curve is the type II excitation energy
for $JN < L \le (J+1)N$.
The angular momentum decreases from $(J+1)N$ to $JN$ according to Eq.~(\ref{TG_stat_am})
along the bold curve.
}\label{fig6}
\end{figure}

In a similar manner to the previous section, we look for the metastable angular-momentum
states by imposing the extremization condition $\partial \tilde{E}_{N,L}^{\rm (II)}(\Omega)/\partial L=0$.
By inspection CMR states $L=JN$ are (either ground or excited) metastable states.
This extremization condition gives another metastable angular momentum,
\bq\label{TG_stat_am}
\bar{L}=(N+1)\left(J+\frac{1}{2}\right)-\Omega,
\eq
and the corresponding energy,
\bq
\label{TG_stat_ene}
\tilde{E}_{N,\bar{L}}^{\rm(II)}(\Omega)
=(N+1)\left[\Omega-\left(J+\frac{1}{2}\right)\right]^2+\frac{N(N+1)}{4}.
\eq
As a function of $\Omega$ this is a parabolic curve the minimum of which is located at
$\Omega=J+1/2$ with energy $N(N+1)/4$, as shown in Fig.~\ref{fig6}. 
Let us next compare this curve with those of
two CMR states $\tilde{E}_{N,L=JN}(\Omega)$ and $\tilde{E}_{N,L=(J+1)N}(\Omega)$.
These CMR energies intersect at $\Omega=J+1/2$ with the energy
$\tilde{E}_{N,JN}(J+1/2)=\tilde{E}_{N,(J+1)N}(J+1/2)=N/4$.
The minimum of~(\ref{TG_stat_ene}) is thus higher than the value $\tilde{E}_{N,JN}$
by $N^2/4$ at $\Omega=J+1/2$.

Another important point is the emergence of certain {\it critical angular frequencies}
where the metastable type II branch disappears and merges into the CMR branch:
\bq
\Omega_{\mathrm{cr}}^{\mp}=\left(J+\frac{1}{2}\right)\mp \frac{N}{2}.
\eq
The stable angular momentum approaches $\bar{L}=(J+1)N$ at $\Omega_{\mathrm{ cr}}^{-}$,
and $\bar{L}=JN$ at $\Omega_{\mathrm{cr}}^{+}$, respectively, and the corresponding energy
coincides with the energy of CMR states.

This is reminiscent of the uniform superflow to dark soliton transition in the weakly interacting limit,
where there exists a critical angular frequency at which the soliton branch bifurcates
from the superflow branch.  Thus a continuous crossover between these topologically
distinct states exists in the TG limit, as well as the weakly interacting limit.
Naturally, we can associate the hole excitations in the TG limit with the soliton branch
in the weakly interacting limit.

\section{Finite-Size Bethe Ansatz}\label{BA}

We studied metastable states drawn from the yrast spectrum in the limits of the weakly and
strongly interacting TG regimes by the extremization condition of the eigenvalues and
by introducing the particle and hole excitations.
The goal of this section is (i) to properly interpolate these limits via the finite-size Bethe ansatz
approach with the hint that the physical properties of the LLH with repulsive interaction are continuous 
and (ii) to show that two distinct phases exist over the entire range of repulsive interactions.
We also vindicate our hypothesis that the soliton branch in the weakly interacting regime are
``hole'' excitations of the quasimomenta in general.
The ground- and excited-state energies in the thermodynamic limit were given by the integral
equation as a continuous limit of the Bethe equations~\cite{63:LL,63:L}.
Recently, the spectrum of LLH was obtained~\cite{06:CB} by treating the
inverse of the TG parameter, which is infinite at the TG regime, as the expansion
parameter, and its analytical interpolation was given~\cite{09:CB}.


\subsection{Bethe equations for quasi-momenta}

We first present how to derive the eigensolutions of the LLH by the Bethe ansatz~\cite{97:KY}.
The delta-function interaction imposes the conditions
\bq
\label{eqn:bethe1}
&&\left[\left( \frac{\partial}{\partial \theta_j}-\frac{\partial}{\partial \theta_k}\right)\Psi\right]_{\theta_j=\theta_k^+}
-\left[\left( \frac{\partial}{\partial \theta_j}-\frac{\partial}{\partial \theta_k}\right)\Psi\right]_{\theta_j=\theta_k^-}
\nonumber\\
&&=\left.g_{\rm 1D}\Psi\right|_{\theta_j=\theta_k}\,.
\eq
Equation~(\ref{eqn:bethe1}) is rewritten by the interchange of the subscripts $j, k$ as
\bq\label{PBC}
\left[\left(\frac{\partial}{\partial \theta_j}-\frac{\partial}{\partial \theta_k}\right)\Psi\right]_{\theta_j=\theta_k^+}
=\left.\frac{1}{2}g_{\rm 1D}\Psi\right|_{\theta_j=\theta_k}\,,
\eq
where $\Psi = \Psi(\{\theta\})$ is the many-body wave function.
The periodic boundary condition is expressed as $\Psi(0,\theta_2,\dots,\theta_N)=\Psi(\theta_2,\dots,\theta_N,2\pi)$.
The entire coordinate space is expressed as
${\cal R}: 0 \le \theta_{{\cal R}_1} < \theta_{{\cal R}_2} < \dots < \theta_{{\cal R}_N} \le 2\pi$ where
$\{{\cal R}_1,{\cal R}_2,\dots,{\cal R}_N\}$ is given by a permutation of $\{1,2,\dots,N\}$.
The next procedure for solving the problem is to restrict the original coordinate space ${\cal R}$
to the ordered space ${\cal R}_I$ and solve the Hamiltonian within the ordered space, say,
${\cal R}_I: 0 \le \theta_1 \le \dots \le \theta_N \le 2\pi$.
The LLH and the condition Eq.~(\ref{PBC}) yield
\bq\label{free}
-\sum_{j=1}^N \frac{\partial^2}{\partial \theta_j^2}\tilde{\Psi}=E\tilde{\Psi} 
\eq
and
\bq\label{R1PBC-1}
\left[\left(\frac{\partial}{\partial \theta_{j+1}}
-\frac{\partial}{\partial \theta_j}\right)\tilde{\Psi}\right]_{\theta_{j+1}=\theta_j}
=\left.\frac{1}{2}g_{\rm 1D}\tilde{\Psi}\right|_{\theta_{j+1}=\theta_j},
\eq
respectively, where $\tilde{\Psi}$ refers to the many-body wave function
in the ordered coordinate space.
The periodic boundary condition and its derivative in the region ${\cal R}_I$ are given by
\bq
\tilde{\Psi}(0,\theta_2,\dots,\theta_N)&=&\tilde{\Psi}(2\pi,\theta_2,\dots,\theta_N)\nonumber\\
&=&\tilde{\Psi}(\theta_2,\dots,\theta_N,2\pi)\,,\nonumber\\\label{R1PBC-2}\\
\left.\frac{\partial}{\partial\theta}\tilde{\Psi}(\theta,\theta_2,\dots,\theta_N)\right|_{\theta=0}
&=&\left.\frac{\partial}{\partial\theta}\tilde{\Psi}(\theta_2,\theta_3,\dots,\theta_N,\theta)\right|_{\theta=2\pi}.\nonumber\\
\label{R1PBC-3}
\eq

The Schr\"{o}dinger equation~(\ref{free}) describes free particles.
All eigenstates and spectra can therefore be represented formally by those of
free particles.
The eigenfunction in the region ${\cal R}$ can be written as
\bq\label{BAwf}
\tilde{\Psi}=\sum_{\cal P}A({\cal P},{\cal R})\exp
\left(i\sum_{j=1}^N \ell_{{\cal P}_j}\theta_{{\cal R}_j}\right),
\eq
where $\{\ell_n\}$ are called {\it quasi-momenta} (or quasi-angular momenta).
This is the basic idea of the Bethe ansatz: one writes down the many-body wave function
in terms of a symmetrized superposition of plane waves with quasi-momenta,
which implicitly includes all the effects of interactions.
This wave function is a superposition of plane waves with $N$ distinct
quasi-momenta $\ell_1 < \ell_2 < \dots < \ell_N$,
${\cal P}$ means $N!$ permutations of quasi-momentum indices, and
$A({\cal P},{\cal R})$ is the coefficient of superposition of plane waves
with a different configuration of quasi-momenta $\{\ell_n\}$.

Substituting the wave function~(\ref{BAwf}) into the conditions (\ref{R1PBC-1}) and (\ref{R1PBC-2}),
we obtain the equations that determine the values of quasi-momenta $\{ \ell_n \}$:
\bq
(-1)^Ne^{-i2\pi \ell_j}=\exp\left[i\sum_{k=1}^N\Theta_{kj}\right],\quad j\in\{1,\dots,N\},
\eq
where
\bq
\Theta_{kj}\equiv -2 \,{\rm arctan}\left[\frac{2(\ell_k-\ell_j)}{g_{\rm 1D}}\right]
\eq
is the two-body phase shift.

Note that the quasi-angular momenta $\{\ell_n\}$ do not have a physical meaning
per se; however, the sum of quasi-angular momenta does
have a physical meaning as the total angular momentum,
\bq\label{bethe_am}
L=\sum_{n=1}^N \ell_n.
\eq
Energy is also given in terms of $\{\ell_n\}$ as
\bq\label{bethe_ene}
E_{N,L}(\Omega=0)=\sum_{n=1}^N \ell_n^{2},
\eq
where the units of quasi-angular momentum and energy are the same as the previous sections,
$\hbar$ and $\hbar^2/(2mR^2)$, respectively.
From the boundary condition we obtain $N$ simultaneous nonlinear equations (Bethe equations),
\bq\label{bethe_eq}
(-1)^{N-1}e^{-2\pi i \ell_n}=
\prod_{m=1}^N \frac{\ell_n-\ell_m
+i g_{\rm 1D}/2}{\ell_n-\ell_m- i g_{\rm 1D}/2},
\eq
which determine the set of values $\{ \ell_n \}$ for each atom $n\in\{1,\ldots,N\}$.
Since all the quasi-momenta are known to be real and continuous for positive $g_{\rm 1D}$,
Eq.~(\ref{bethe_eq}) can be separated into real and imaginary parts, both of which
are found to give the same set of solutions.  Therefore it is sufficient to solve only
the real part of the set of equations.


\subsection{Weakly interacting regime}

\begin{figure}
\includegraphics[scale=0.4]{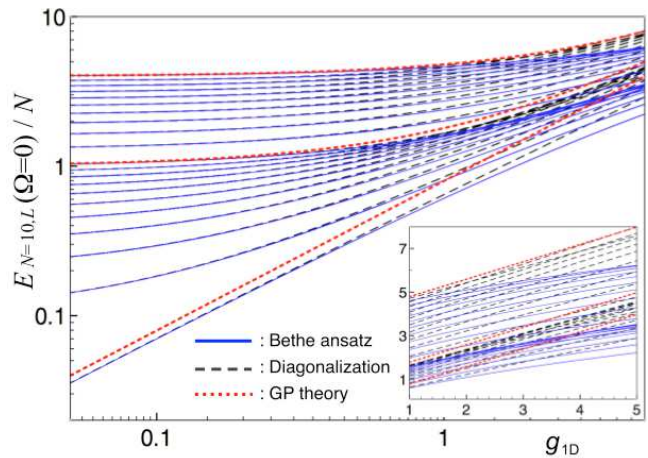}
\caption{
(Color online) Comparison of exact yrast eigenenergies of Eq.~(\ref{bethe_ene}) obtained from
the Bethe ansatz, yrast states obtained from the diagonalization of
Hamiltonian~(\ref{Ham2}) for $N=10$, and the plane-wave energy branches
for $J\in\{0,1,2\}$ given by the mean-field theory.
For Bethe ansatz and diagonalization results, the ground ($L=0$), and
$2N$ yrast states $|L|\in\{1,\ldots, 2N\}$ are plotted on a log scale.
The inset enlarges the region $1\le g_{\rm 1D} \le 5$ on a linear scale.
}\label{fig7}
\end{figure}

We numerically solve the real part of the Bethe equations~(\ref{bethe_eq}) for
each set of energy levels
characterized by the different total angular momenta.
The numerical solution of Eqs.~(\ref{bethe_eq}) is highly sensitive to the initial
set of trial values of $\{\ell_n\}$.
If this initial set is sufficiently close to a solution for a target angular-momentum state,
the set of solutions $\{\ell_n\}$ can be correctly obtained.  In contrast, if the initial set
is closer to another angular-momentum state, the total angular momentum given by Eq.~(\ref{bethe_am})
reveals undesired jumps, deviating from the target angular momentum. In such a case we again start from
another initial set of trial values of quasi-momenta.

For simplicity we start with consideration of the trivial noninteracting limit $g_{\rm 1D}=0$ where all
the quasi-angular momenta $\{\ell_n\}$, and hence the total angular momentum $L$, are zero for the ground state.
The energy of the first excited state corresponds to the degenerate yrast levels
$E_{N,L=\pm 1}(\Omega=0)$,
where only one of the quasi-angular momenta has the value $1$ or $-1$ and the
remaining quasi-angular momenta are zero.
The second excited state has total angular momentum $L=\pm 2$, that is,
two of the quasi-angular momenta take the value $|\ell_n|=1$.
Higher excited states can be obtained in a similar way.
Starting from these initial conditions, solutions can be obtained
from the non-interacting to the strongly interacting regime by gradually increasing
the value of $g_{\rm 1D}$.

Results of the Bethe ansatz are obtained by the following steps:
\begin{enumerate}

\item For $L\in\{0,1,\ldots, \lfloor N/2\rfloor\}$, the quasi-momenta are given by
$\{\ell_n\}\in
\{(0,\ldots, 0)$,
$(1,0,\ldots,0)$,
$(1,1,0,\ldots, 0)$, $\ldots$\,,
$(1,1,1,1,1,0,0,0,0,0)\}$
for $g_{\rm1D}=0$ and for $N=10$, for example.
Each set is directly calculated by solving Eqs.~(\ref{bethe_eq})
with the target angular-momentum state from $L=0$ to $L=N/2$, respectively.

\item The $L\in\{\lfloor N/2\rfloor,\lfloor N/2\rfloor+1,\ldots, N\}$ states are obtained from a transformation of
the first $\lfloor N/2\rfloor-1$ states of (i) via $L=N-\tilde{L}$, where
$\tilde{L}\in\{\lfloor N/2\rfloor,\lfloor N/2\rfloor-1,\ldots, 0\}$.
The quasi-momenta are given by a transformation of the following form.  For example,
$\tilde{L}=1$ with $\{\ell_n\}=(1,0,0,0,0,0,0,0,0,0)$ transforms to $L=9$ with
$\{\ell_n\}=(0,1,1,1,1,1,1,1,1,1)$.

\item There exist degenerate spectra with the same magnitude of angular momentum
$E_{L}=E_{-L}$ for the yrast states.
The corresponding quasi-momenta are given by a transformation,
$L: (\ell_1,\ell_2,\ldots,\ell_N)\to -L: (-\ell_1, -\ell_2,\ldots,-\ell_N)$.
Although there exist a denumerably infinite number of other kinds of excitations of
higher energy, these states are irrelevant for comparison with the previous results.
Hence we do not use the larger quantum-number subscript $q \ge 2$,
as in previous sections.

\item We gradually increase $g_{\rm 1D}$ from zero with the step size
$O(10^{-4})$ for $g_{\rm 1D} \lesssim O(1)$,
$O(10^{-2})$ for $O(1) \lesssim g_{\rm 1D}\lesssim O(10)$, and
$O(10^{-1})$ for $O(10) \lesssim g_{\rm 1D}\lesssim O(10^2)$.
The convergence of the numerical solutions of Eqs.~(\ref{bethe_eq}) is confirmed by comparing
both sides of the Bethe equation with the substitution of the solution $\{\ell_n\}$.
We set the tolerance factor, i.e., the difference in the left- and right-hand sides of Eqs.~(\ref{bethe_eq}),
to be $10^{-8}$. We also required as a secondary convergence criterion that the total angular momentum
be conserved to better than $10^{-8}$ in the target angular momentum.
If, during the changing of $g_{\rm 1D}$, the angular momentum
unexpectedly deviates from the target angular momentum, and/or
some of quasi-angular momenta show a jump as a function of $g_{\rm 1D}$, these errors are 
detectable. For the attractive case, complex solutions of Eqs.~(\ref{bethe_eq}) appear~\cite{07:SDD},
indicating ground-state soliton formation, which we do not treat here.
\end{enumerate}

In Fig.~\ref{fig7}, we compare low-lying excited
states obtained by three different theoretical methods:
Bethe ansatz, diagonalization, and GP mean-field theory.
We note that the concept of yrast state for the angular momenta $JN < L < (J+1)N$
does not exist in the mean-field theory: this theory is concerned only with
the single-particle angular momentum, which coincides with the average angular momentum
in this theory. We thus plot the mean-field energy for the integral single-particle angular momenta.
We plot the first $4N+1$ yrast spectra $E_{N,L}$, $|L|\in \{0,\dots,2N\}$,
as a function of the strength of interaction $g_{\rm 1D}$ for $N=10$.
As expected, the rigorous Bethe ansatz spectra have the lowest energy for any $g_{\rm 1D}$,
the mean-field plane-wave branch has the highest value, and diagonalization results
have values in between those obtained by the Bethe ansatz and mean-field theory.
For $g_{\rm 1D} \lesssim 1$, the Bethe ansatz and diagonalization results quantitatively
agree very well, while the latter becomes larger than the exact spectra for $g_{\rm 1D} \gtrsim 1$~\cite{note2}.


\subsection{Medium- to strongly interacting regime}

\begin{figure}
\includegraphics[scale=0.45]{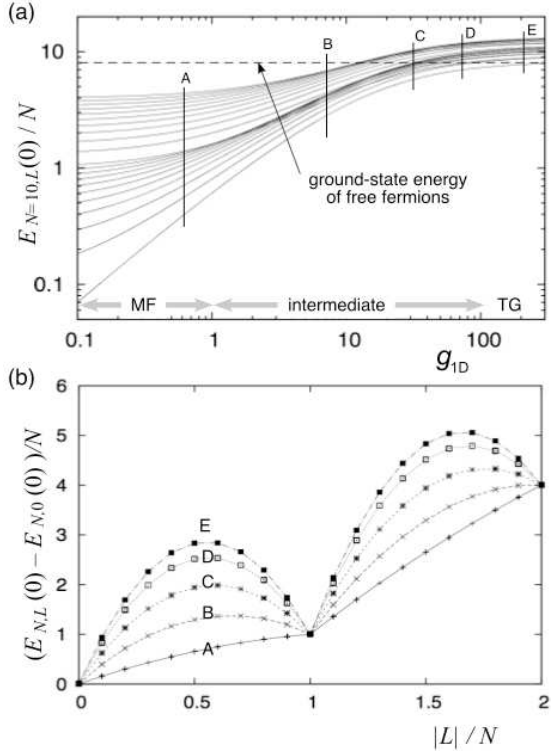}
\caption{
(a) Eigenvalues $E_{N,L}$ of the Lieb-Liniger Hamiltonian~(\ref{LLH})
obtained by the finite-size Bethe ansatz as a function of interaction $g_{\rm 1D}$ for
$L=0$ (ground state), and $L\in\{\pm 1, \ldots, \pm 2N\}$ (excited states)
where $E_L=E_{-L}$. The horizontal dashed line corresponds to the energy of
free fermions (\ref{TG_gs}) with the same number of atoms.
(b) Excitation energies $(E_{N,L}-E_{N,0})/N$ of yrast states as a function
of angular momentum $|L|/N$ for fixed strengths of interaction.
A, B, C, D, and E correspond to the vertical line in (a).
At $L=JN\in \{0,\pm N,\pm 2N,\ldots\}$
the energy is that of a CMR state (relative to the unboosted ground state), therefore independent of
$g_{\rm 1D}$, and is given by $J^2$, even in the thermodynamic limit.
}\label{fig8}
\end{figure}

\begin{figure}
\includegraphics[scale=0.43]{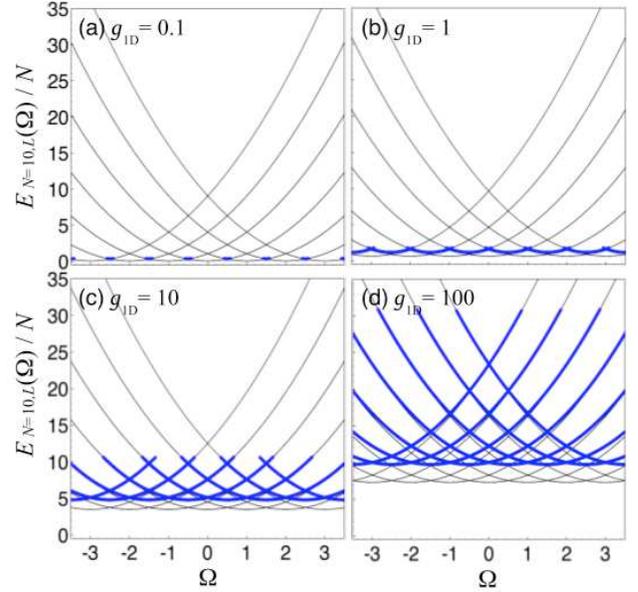}
\caption{
(Color online) Eigenvalues of the Hamiltonian~(\ref{rotH}) that satisfy the extremization condition 
$\partial E_{L,N}/\partial L=0$ for (a) weakly interacting mean-field regime,
(b) and (c) medium-interaction regime, and (d) strongly interacting TG regime.
The corresponding average angular momentum of the thin curves is given
by integers $L/N=J$.
The thick curves, whose average angular momentum is noninteger, come
from the type II excitation branch.
}\label{fig9}
\end{figure}

\begin{figure}
\includegraphics[scale=0.7]{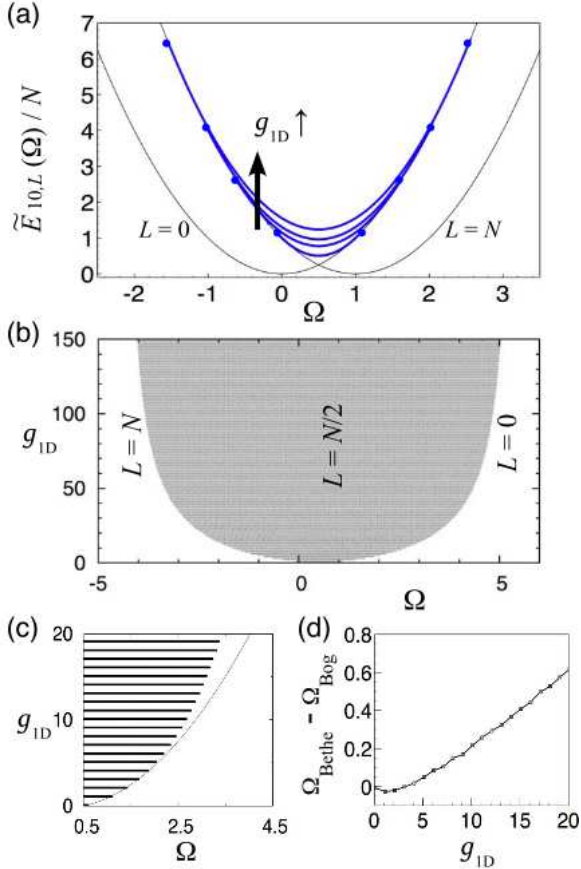}
\caption{
(a) Two CMR states with $L=0,N$ (thin curves) and
the type II branch (thick curve) that connects them.
The energy is defined relative to the interaction energy $V_{\rm int}$.
(b) The existence range of the type II branch with angular momenta $L\in [0,N]$.
At the left (right) boundary, the angular momentum is $L=N$ ($L=0$),
and $L=N/2$ at $\Omega=0.5$ (vertical dashed line), decreasing
linearly with respect to $\Omega$ in the shaded area.
(c) Enlargement of (b) in the weakly interacting regime. The solid curve is
the phase boundary given by the Bogoliubov theory.
(d) Difference between the phase boundaries given by the Bethe ansatz and
the Bogoliubov theory. The Bogoliubov phase boundary overestimates the
extent of the shaded area.
}\label{fig10}
\end{figure}

We now further investigate the yrast states via the Bethe ansatz, going beyond
the weakly interacting mean-field regime.
Figure~\ref{fig8}(a) plots $E_{N,L}(\Omega=0)$ for $|L|\in\{0,\pm 1,\ldots, \pm 2N\}$
over a wide range of repulsive interactions, where the horizontal dotted line shows
the ground-state energy of free fermions given by Eq.~(\ref{TG_gs}).
All the ground- and excited-state energies monotonically increase with respect
to $g_{\rm 1D}>0$.

Note, however, that the energy does not monotonically increase with respect to
the total angular momentum for a fixed strength of interaction. This is clearly shown
in Fig.~\ref{fig8} (b), where we show the excitation energy $(E_{N,L}-E_{N,L=0})/N$ of yrast states
for several values of fixed interaction strengths (indicated as A, B, C, D, and E)
as a function of $L/N$.
In the noninteracting limit, the the yrast spectrum is linear with respect to $|L|/N$
with nodes at $|L|/N \in \mathbb{Z}$.
While in a weakly interacting limit (plot A), the spectrum still looks almost linear,
as the interaction $g_{\rm 1D}$ increases, the kinks in the yrast spectra at
$L=JN\in \{0,\pm 1,\dots\}$ become more pronounced due to the large increase in the energy
of the yrast state in between $JN < L < (J+1)N$. For strong interactions (curve E),
the system is in the TG regime, which can be confirmed by the fact that
the excitation energy has the value between Eq.~(\ref{TG_exc}) for $N=9$ and $N=11$.

Finally, we observe numerically that
the excitation energy of the CMR state $L=JN$ is independent of $g_{\rm 1D}$, and is given
by Eq.~(\ref{sf_ex_ene}), namely $(E_{N,L}-E_{N,0})/N=(L/N)^2$.
This follows from the nature of the CMR state $L=JN$,
which is just a Galilean boost of the nonrotating state;
under this transformation interactions are unchanged.


In order to see how these points are transformed in the rotating frame, we
again rotate the yrast spectrum according to
\bq\label{ene_rot}
E_{N,L}(\Omega)&=&\sum_{j=1}^N (\ell_j-\Omega)^2\nonumber\\
&=& E_{N,L}(0) -2\Omega L +\Omega^2N.
\eq
The results are shown in Fig.~\ref{fig9} for various strengths of interaction.
The thin curves are parabolas $(\Omega-J)^2+V_{\rm int}$ for various values of
center-of-mass quantum numbers $J$.
The lowest possible energy of the CMR state is thus given by $V_{\rm int}$ at $\Omega=J \in \{\mathbb{Z}\}$.

The thick curves plot other metastable angular-momentum states,
the angular momentum of which is given by a nonintegral multiple of $N$.
The weakly interacting mean-field regime is shown in Fig.~\ref{fig9}(a), where
the type II branch that satisfies the metastable condition just starts to appear.
Thus these are the energies of the quantum solitons [see also Fig.~\ref{fig3}(a)].
As the interaction increases [Figs.~\ref{fig9}(b) and \ref{fig9}(c)]
the domain with the swallow-tail shape
enclosed by the two CMR branches, as well as the size of the type II branch, increases.
In the TG limit [\ref{fig9}(d)], the area of the swallow-tail region saturates the spectra.


These behaviors are quantitatively summarized in Fig.~\ref{fig10}(a),
which shows the energy $E_{N,L}(\Omega)/N$ of metastable states relative to
the interaction energy $V_{\rm int}$ at each strength of interaction.
The CMR branches drawn by the thin curves no longer have a $g_{\rm 1D}$-dependence 
because of the subtraction of $V_{\rm int}$,
while the thick curve gradually increases the domain over which it extends as $g_{\rm 1D}$ increases.
For simplicity we plot only two CMR branches with angular momenta $L=0,N$,
and a metastable state associated with the type II branch that smoothly connects these two CMR states.

As $\Omega$ increases, the thick curve appears to bifurcate from the CMR branch
with angular momentum $L=N$ at a certain $\Omega (< 0.5)$, and
at $\Omega=0.5$ the energy becomes minimum.
As $\Omega$ increases further, this branch
smoothly merges into the CMR branch with angular momentum $L=0$ 
and eventually disappears at a certain $\Omega (> 0.5)$.
We therefore find that the same kind of energy bifurcation which
was found in the mean-field theory
persists over the full range of repulsive interactions.

Figure~\ref{fig10}(b) plots the existence range of the metastable-state
type II excitation branch.
The shaded area indicates the existence of such a branch.
For higher angular-momentum uniform superflow states $L=(J+1)N$ and $L=JN$,
the boundary can be obtained by
the parallel displacement of this figure along the $\Omega$ axis.
The angular momentum on the phase boundary $\Omega (< 0.5)$ is given by $L=N$, and
the angular momentum linearly decreases as $\Omega$ increases, just like in Fig.~\ref{fig4} (a).
At $\Omega=0.5$ (vertical dashed line), the value of the angular momentum is
given by $L=N/2$ irrespective of the strength of interaction.
At a certain value of $\Omega (> 0.5)$ the angular momentum eventually
goes to zero, causing the metastable hole excitation branch to disappear.
This behavior corresponds to the fact that in the mean-field theory
the type II branch bifurcates from the plane-wave regime,
developing nodes, and it again merges into the plane-wave regime
with the increase of $\Omega$~\cite{08:KCU, 09:KCU}.
The phase boundary approaches $\Omega=(J+1/2)\pm N/2$ in the
strongly interacting regime.

In the lower panel of Fig.~\ref{fig10} the phase boundary is compared with
the one obtained by Bogoliubov theory in the weakly interacting regime.
Bogoliubov theory thus predicts the quantitatively correct
phase boundary to the 5$\%$ level in the weakly interacting limit $g_{\rm 1D} \lesssim 5$ (for $N=10$),
but it significantly overestimates the phase boundary as the interaction increases.

These results indicate that the continuous change in the topologically distinct quantum
phases can be found at any strength of interaction.  For larger strength of interaction,
the existence range of the type II branch increases.
In this existence range the angular momentum changes linearly in $\Omega$,
and the rate of change thus decreases for larger coupling constant.


\section{Conclusions}\label{conclusion}

We addressed the continuous topological change in the repulsively interacting
1D Bose gas on a ring, previously found in the Gross-Pitaevskii mean-field theory~\cite{09:KCU}.
In the mean-field theory the Gross-Pitaevskii equation has two kinds of solutions: uniform
superflow and the broken-symmetry soliton train as a function of interaction strength and rotation.
In the weakly interacting regime, the energy diagram is characterized by the smooth bifurcation
of a soliton branch from a plane-wave branch in the rotating frame, which is the key to the continuous
change in the topology of the condensate wave function characterized as a self-induced phase slip.

In this article we vindicated this picture starting from the many-body Hamiltonian without assuming
the existence of the condensate wave function and spontaneous symmetry breaking.
We solved the yrast problem of the original Lieb-Liniger Hamiltonian by
three methods: diagonalization of the Hamiltonian in the weakly interacting regime;
Bose-Fermi mapping in the strongly interacting TG regime; and
the Bethe ansatz approach for all regimes of repulsive interaction strength.

We then obtained the eigensolutions in the rotating frame through transforming the eigenvalues
according to specific values of the angular frequency of the external rotating drive $\Omega$.
The extremization condition is imposed so that eigensolutions which are realizable
in practice are extracted from a very large number of possibilities. The realizable states,
namely those metastable under symmetry-breaking perturbations, 
reveals that two kinds of eigensolutions are physically distinguished. One is the superflow state
in which angular momentum is an integral multiple of the number of atoms. 
The other is a \emph{quantum soliton} characterized by a set of soliton components,
which are also the yrast states.
In the weakly interacting regime, the energy and angular momentum obtained by
exact diagonalization and the Bethe ansatz agree well with those predicted by
the Gross-Pitaevskii equation. This fact bears out the above physical meanings of
metastable states.
In the opposite limit, the strongly interacting TG limit was studied by
the Bose-Fermi mapping. We introduced the concept of particle and
hole excitations, which are well defined not only in the fermionized system
but also for the whole interaction range in terms of the quasimomenta.
The solution was similarly transformed into the rotating frame
to extract the metastable states.

In between the weakly interacting mean-field and strongly interacting TG limits,
we employed the Bethe ansatz approach.
In order to compare with the diagonalization results,
the set of Bethe equations was solved without substituting the summation with an integral,
and the lowest $(2N+1)$ discrete excited states were found.
By the same transformation into the rotating frame, we elucidated how the energy diagram
of these topologically distinct states changes as the strength of interaction increases.

Energy and angular momentum of the two kinds of
topologically distinct states exist over the whole of repulsive interaction $g_{\rm 1D}$.
The quantum phase diagram in the $\Omega$-$g_{\rm 1D}$ plane for the quantum soliton
with a single density notch and with angular momentum $0 < L < N$ was explicitly shown.
This metastable quantum phase transition is technically a crossover: it can occur only
in a finite system, as expressed by our choice of units in terms of the ring circumference,
and all states are connected analytically.  Nevertheless, one finds a sharp change between distinct physical states which will appear as a QPT in experiments.

\section*{ACKNOWLEDGMENTS}

We thank Joachim~Brand, Marvin~Girardeau, Ewan~Wright, and Tetsuo~Deguchi for useful discussions.
This work was supported by the National Science Foundation under Grant PHY-0547845
as part of the NSF CAREER program (L.D.C.), a Grant-in-Aid for Scientific Research under Grant Numbers 17071005 (M.U.) and 21710098 (R.K.), and by the Aspen Center for Physics.



\begin{thebibliography}{50}
\bibitem{note2}
The convergence is better for (i) a larger number of atoms,
(ii) lower eigenvalues, and (iii) smaller strength of interaction.
Figure 7 indicates that for $N=10$ and up to the yrast states $|L|=2N$,
the deviation of the exact diagonalization results from the
Bethe ansatz results starts to emerge for $g_{\rm 1D} \gtrsim 1 $.
The results for $N=40$ in this section has, in general, better
agreement with the rigorous results.
\end{thebibliography}
\end{document}